\documentclass[twocolumns]{aa}  
\usepackage{graphicx}
\usepackage{txfonts}
\usepackage{natbib,twoopt}
\bibpunct{(}{)}{;}{a}{}{,}
\usepackage{graphicx}
\usepackage{txfonts}
\usepackage{hyperref}
\hypersetup{
    colorlinks=true,
    linkcolor=blue,
    citecolor = blue,
    filecolor=magenta,      
    urlcolor=blue}
\graphicspath{{./}{figures/}}
\begin{document} 

\authorrunning{Billi et al.}
\titlerunning{Rotational velocities of Blue Straggler Stars in the Globular Cluster M55}

\title{Rotational velocities of Blue Straggler Stars in the Globular Cluster M55\thanks{Based on observations collected at the European Southern Observatory, Cerro Paranal (Chile), under Program 093.D-0270 (PI: Lovisi).}}

\author{A. Billi\inst{1}\fnmsep\inst{2},
    F. R. Ferraro\inst{1}\fnmsep\inst{2},
    A. Mucciarelli\inst{1}\fnmsep\inst{2},
    B. Lanzoni\inst{1}\fnmsep\inst{2},
    M. Cadelano\inst{1}\fnmsep\inst{2},
    and L. Monaco\inst{3} }

\institute{Dipartimento di Fisica e Astronomia, Università degli Studi di Bologna, Via Gobetti 93/2, I-40129 Bologna, Italy\\
\email{alex.billi2@unibo.it}
\and
INAF, Osservatorio di Astrofisica e Scienza dello Spazio di Bologna, Via Gobetti 93/3, I-40129 Bologna, Italy
\and
Universidad Andres Bello, Facultad de Ciencias Exactas, Departamento de Ciencias Físicas - Instituto de Astrofisica, Autopista Concepción-Talcahuano, 7100, Talcahuano, Chile
}



\abstract{By using high-resolution spectra acquired with
  FLAMES-GIRAFFE at the ESO/VLT, we measured radial and rotational
  velocities of 115 stars in the Galactic globular cluster M55. After
  field decontamination based on the radial velocity values, the final
  sample of member stars is composed of 32 blue straggler stars (BSSs)
  and 76 reference stars populating the red giant and horizontal
  branches of the cluster. In agreement with previous findings, the
  totality of red giant branch stars has negligible rotation ($<$ 10
  km s$^{-1}$), and horizontal branch stars have rotational velocities
  of 40 km s$^{-1}$ at most. In contrast, the BSS rotational velocity
  distribution shows a long tail extending up to $\sim$ 200 km
  s$^{-1}$, with 15 BSSs (out of 32) spinning faster than 40 km
  s$^{-1}$. By defining the threshold for fast rotating BSSs at 40 km
  s$^{-1}$, this sets the percentage of these stars at 47 $\pm$ 14 \%. Such a
  large value has never been found before in any globular clusters. It is roughly comparable to that measured in other loose systems ($\omega$
  Centauri, M4, and NGC 3201) and significantly larger than that
  observed in high-density clusters (as 47 Tucanae, NGC 6397, NGC
  6752, and M30). This evidence supports a scenario where recent BSS
  formation is occurring in low-density environments. We also find
  that the BSS rotational velocity tends to decrease for decreasing
  luminosity, as found for another loose cluster of the sample
  (namely, NGC 3201).}

\maketitle
%

\section{Introduction}
\label{sec:intro}
Blue Straggler Stars (BSSs) are an ``exotic'' stellar population, not
predicted by the standard stellar evolution theory. In the
color-magnitude diagram (CMD) of Galactic globular clusters (GCs), they lie on a bluer
(hotter) and brighter extension of the main sequence (MS; see, e.g.,
\citealt{ferraro+92, ferraro+97, ferraro+01, ferraro+03}), where stars
more massive (M $\sim$ 1.2-1.6 M$_{\odot}$; \citealt{Shara_97};
\citealt{Gilliland_98}; \citealt{DeMarco_05}; \citealt{Fiorentino_14};
\citealt{Raso_19}) than normal MS turnoff stars (with M $\sim 0.8$
M$_\odot$) spend their core hydrogen-burning phase. Due to their large
mass, BSSs are crucial gravitational test particles to probe internal
dynamical processes occurring in GCs. \citealt{Ferraro_12} proposed the concept of ``dynamical clock'',
a method based on the analysis of the BSS radial distribution to infer
the level of dynamical evolution of star clusters. In this respect,
the $A^{+}_{rh}$ parameter (\citealt{Alessandrini_16, Lanzoni_16})
measures the degree of BSS central segregation with respect to a
lighter reference population (typically red giant branch or MS stars)
adopted as reference. This parameter has been found to efficiently
trace the level of dynamical evolution experienced by star clusters
(\citealt{Ferraro_18, ferraro+19, ferraro+20, Ferraro_23a}).

Two main scenarios for BSS formation have been suggested so far: mass
transfer (MT) activity in binary systems \citep{McCrea_64}, where the
companion star transfers mass and angular momentum to the accreting
proto-BSS that then becomes a MT-BSS, and direct collisions between
two or more stars, producing collisional BSSs (COL-BSSs;
\citealt{HillsDay_76}).
The frequency of both formation channels can in principle be 
increased by mergers/collisions produced by the evolution of triple systems
(\citealt{Andronov_06, PeretsFabrycky_09}). To shed light on the
formation and the evolutionary processes of this class of exotica,
several studies have been carried out in both open and globular
clusters searching for evidence of the two formation channels and
possible links between the observational properties of BSSs and those
of the host parent cluster (e.g., \citealt{ferraro+93, ferraro+95,
  davies+04, piotto+04, Sollima_08, MathieuGeller_09, geller+11,
  Gosnell_14, leigh+07, leigh+13}).

Since MT-BSSs are produced by binary systems, this formation channel
should be favored in low-density environments because, in conditions
of high stellar crowding, binary systems are subject to multiple
dynamical interactions that can disrupt them.  Indeed, promising
evidence of MT-BSSs has been found from both photometric and
spectroscopic studies in low-density environments, such as loose GCs
and open clusters.  By analyzing \textit{Hubble} Space Telescope (HST)
photometric data of a sample of loose Galactic GCs, \citealt{Sollima_08}
found a positive correlation between the fraction of binary systems
and that of BSSs, in agreement with what expected if MT activity in
binaries is the main BSS formation channel.  Moreover, an ultraviolet
excess has been detected (\citealt{Gosnell_14, Gosnell_15}) in a few
BSSs in the open cluster NGC 188, a stellar system with a very large
fraction of BSSs in binary systems (\citealt{MathieuGeller_09}). This has been
interpreted as the evidence of the presence of a hot white dwarf (the
remnant core of the original donor star at the end of the MT phase),
orbiting a newly formed BSS. From the chemical point of view,
MT-BSSs are predicted to show anomalous surface abundances of carbon,
nitrogen and oxygen, due to material partially processed by the CNO
burning cycle in the innermost layers of the donor star, that is now
accumulated onto the BSS surface
(\citealt{SarnaDeGreve_96}). Conversely, COL-BSSs are expected to show
normal chemical abundances (\citealt{Lombardi_95}).  A sub-sample of
BSSs with carbon and oxygen depletion (the expected MT formation
signature) has been indeed found both in 47 Tucanae
(\citealt{Ferraro_06}) and in M30 (\citealt{Lovisi_13a}). Similarly, an enhancement in barium (\citealt{Milliman_15}, \citealt{Nine_24}) has been found in a few BSS in open clusters and interpreted as a signature of mass transfer from an asymptotic giant branch companion onto a proto-BSS.

A potential manifestation of COL-BSS comes from the photometric
investigation of post-core collapse GCs. In fact, during the core
collapse phase, the central density increases significantly, and this
can trigger an enhancement of the collisional activity that
contributes to the formation of COL-BSSs.  Thus, the presence of a
relatively new and coeval population of COL-BSSs with different masses
is expected to be observable in stellar systems that experienced core
collapse in their recent past.  In accordance with this, the narrow
blue sequence of BSSs, which appears to be well separated from a
redder BSS sequence in the CMD of a few post-core collapse GCs (namely
M30, NGC 362, M15 and NGC 6256; see \citealt{Ferraro_09,
  Dalessandro_13, Beccari_19, Cadelano_22}) has been interpreted as due
to a population of COL-BSSs recently formed during the core collapse
event.

The rotational velocity, instead, seems to not allow distinguishing
between MT- and COL-BSSs.  In fact, large rotational velocities are
expected at birth for BSSs formed by both the channels, due to the
transfer of mass and angular momentum for MT-BSSs
(\citealt{Packet_81}; \citealt{SarnaDeGreve_96}; \citealt{deMink_13}),
and to the conservation of angular momentum for COL-BSSs
(\citealt{BenzHills_87, sills+02}).  However, braking mechanisms (such
as magnetic braking and disk locking) are predicted to slow down the
rotation of these stars as function of time (\citealt{LeonardLivio_95,
  Sills_05}; \citealt{Leiner_18}, \citealt{Sun_24}). The characteristic timescales of these processes are
still largely unknown, and any new observational constraint is
precious in this respect.  Thus, although the rotational velocity
cannot be used as a diagnostic of the formation mechanism, it could be
an indicator of the BSS age (i.e., the amount of time passed from the
end of MT, or from the collision that originated the BSS), and provide
new physical information for a proper modeling of these objects.

In this context, several years ago our group started an
high-resolution spectroscopic survey aimed at the chemical and
kinematical characterization of BSSs belonging to GCs with different
structural properties. Eight GCs have been observed so far: 47 Tucanae
(\citealt{Ferraro_06}), NGC 6397 (\citealt{Lovisi_12}), NGC 6752
(\citealt{Lovisi_13b}), M30 (\citealt{Lovisi_13a}), M4
(\citealt{Lovisi_10}), $\omega$ Centauri (\citealt{Mucciarelli_14}),
NGC 3201 (\citealt{Billi_23}), and M55 (this paper). Using this
dataset, \citealt{Ferraro_23b} found intriguing relations between the
fraction of fast-rotating (FR) BSSs and the parent cluster structural
parameters (such as central density and concentration), suggesting
that rapidly spinning BSSs preferentially populate low-density
environments. According to \citealt{Ferraro_23b}, 
FR-BSSs have been defined as those with projected
rotational velocity v sin(i) $\geq$ 40 km s$^{-1}$, with $i$ being the
inclination angle on the plane of the sky. In that paper, we
analyzed the overall distribution of rotational velocities 
in a sample of 300 BSSs in 8 clusters, finding that it shows a 
significant drop at values larger than 30-50 km/s, thus suggesting an 
appropriate threshold value of 40 km/s.  We also showed that 
the assumption of a slightly different threshold value does not affect the results.
Among the investigated
clusters, M55 is the one with the lowest values of central density
($\log\rho_0=2.2$ in units of L$_\odot$ pc$^{-3}$) and concentration
parameter ($c=0.93$). This paper is devoted to the detailed presentation and
discussion of the results obtained for the BSSs and the reference
horizontal branch (HB) and red giant branch (RGB) stars observed in
M55.  We organized the paper in the following sections: the
observations and data reduction are discussed in Section
\ref{sec:obs}; the radial velocities measures and the determination of
the atmospheric parameters for the cluster member stars are described
in Sections \ref{sec:rv_membership} and
\ref{sec:atmospheric_parameters}, respectively; the determination of
rotational velocities is presented in \ref{sec:rotational_velocities};
Section \ref{sec:discussion} describes the obtained results and our
conclusions.

\begin{table*}
\caption{Properties of the analyzed BSSs.}
 \label{table:1} 
    \centering 
    \begin{tabular}{c c c c c c c c} 
        \hline\hline 
        ID & R.A. & Decl. & B & (B$-$I) & V$_R$ & v sin(i) & Var \\ 
           & [deg] & [deg] & & & [km s$^{-1}$] & [km s$^{-1}$] & \\
        \hline 
        100108 & 294.9850000 & -30.9720000 & 15.99 & 0.29 & 178.3 $\pm$ 0.2 & $>$ 200 & \\ 
        100101 & 294.9927917 & -30.9853056 & 16.21 & 1.04 & 178.1 $\pm$ 1.6 & 21 $\pm$ 3 & SXP \\  
        2700607 & 295.0269583 & -30.9911389 & 16.25 & 0.68 & 177.5 $\pm$ 8.5 & 107 $\pm$ 15 & SXP \\
        200142 & 295.0122083 & -30.9582222 & 16.35 & 0.66 & 179.3 $\pm$ 1.9 & 95 $\pm$ 6 & SXP  \\
        2700663 & 294.9460417 & -30.9597222 & 16.54 & 0.98 & 174.8 $\pm$ 1.9 & 20 $\pm$ 4 & SXP \\
        2700733 & 295.0367500 & -30.9546111 & 16.59 & 0.60 & 147.8 $\pm$ 12.0 & 202 $\pm$ 15 &	SXP\\
        2800072 & 295.0967083 & -30.9473611 & 16.59 & 0.96 & 173.6 $\pm$ 1.3 & 125 $\pm$ 4 &	\\
        2700739 & 294.9941667 & -30.9064444 & 16.63 & 0.68 & 169.4 $\pm$ 4.9 & 69 $\pm$ 4 & \\
        2700785 & 294.9553750 & -30.9422222 & 16.73 & 0.77 & 163.3 $\pm$ 0.1 & 68	$\pm$ 5 & SXP \\
        2700747 & 294.9772917 & -30.9997222 & 16.76 & 0.83 & 174.5 $\pm$ 6.7 & 70 $\pm$ 4 & SXP \\
        2700813 & 295.0688333 & -30.9846944 & 16.79 & 0.64 & 186.7 $\pm$ 2.9 & 39	$\pm$ 3 & \\
        2700834 & 295.0217500 & -30.9895556 & 16.84 & 0.59 & 182.1 $\pm$ 8.9 & 104 $\pm$ 10 & EW \\
        200178 & 295.0122917 & -30.9748333 & 16.85 & 0.74 & 179.6 $\pm$ 1.6 & 17 $\pm$ 2 & SXP \\
        200196 & 294.9982500 & -30.9483611 & 16.89 & 0.60 & 176.8 $\pm$ 1.3 & 225 $\pm$ 15 & SXP \\
        200204 & 295.0167500 & -30.9705278 & 16.91 & 0.77 & 173.4 $\pm$ 4.5 & 199	$\pm$ 15 & SXP \\
        2700918 & 294.9522917 & -30.9461389 & 17.03 & 0.72 & 173.0 $\pm$ 1.1 & 62 $\pm$ 10 & SXP \\
        2700925 & 294.9459583 & -30.8815278 & 17.03 & 0.67 & 179.3 $\pm$ 1.4 & 17 $\pm$ 3 & \\
        100215 & 294.9859583 & -30.9600833 & 17.05 & 0.50 & 182.0 $\pm$ 6.2 & 19 $\pm$ 2 & \\
        2701036 & 295.0175000 & -30.9150278 & 17.23 & 0.89 & 176.6 $\pm$ 1.4 & 37 $\pm$ 2 & \\
        200256 & 295.0286250 & -30.9425556 & 17.32 & 0.78 & 177.8 $\pm$ 1.2 & 12 $\pm$ 2 & SXP \\
        2701130 & 295.0383333 & -30.9453056 & 17.32 & 0.89 & 167.5 $\pm$ 0.7 & 9 $\pm$ 3 & SXP \\
        100252 & 294.9789583 & -30.9728611 & 17.35 & 0.34 & 176.1 $\pm$ 3.6 & 4 $\pm$ 1 & SXP \\
        2701162 & 294.9395000 & -30.9343611 & 17.39 & 0.79 & 175.7 $\pm$ 4.0 & 40 $\pm$ 4 & SXP \\
        2701145 & 294.9740000 & -31.0132222 & 17.39 & 0.81 & 172.5 $\pm$ 1.2 & $<$ 3 & SXP \\
        100256 & 294.9752083 & -30.9690278 & 17.46 & 0.81 & 180.3 $\pm$ 5.6 & 35 $\pm$ 4 & SXP \\
        100280 & 294.9942083 & -30.9569444 & 17.47 & 0.67 & 174.3 $\pm$ 3.5 & $<$ 3 & SXP \\
        2701235 & 295.0499583 & -31.0348889 & 17.50 & 0.72 & 174.2 $\pm$ 3.4 & 26 $\pm$ 3 & SXP \\
        2701308 & 295.0078750 & -30.9275556 & 17.53 & 0.76 & 171.3 $\pm$ 4.1 & 69 $\pm$ 6 & SXP \\
        2701374 & 294.9658750 & -30.9316389 & 17.57 & 0.81 & 189.3 $\pm$ 1.1 & $<$ 3 & SXP \\
        2701376 & 295.0330417 & -30.9474167 & 17.63 & 0.93 & 189.1 $\pm$ 1.4 & 59 $\pm$ 4 & SXP \\
        200354 & 295.0214583 & -30.9805278 & 17.68 & 0.88 & 174.2 $\pm$ 1.4 & 20 $\pm$ 4 & \\
        2701676 & 294.9928333 & -30.9187778 & 17.92 & 1.08 & 171.2 $\pm$ 1.6 & $<$ 3 & \\ 
        \hline 
    \end{tabular}
    \tablefoot{Identification number, coordinates, B-band magnitude and (B$-$I)
  color (from \citealt{Stetson_19}), radial velocity, rotational velocity, and variability type
  of the analyzed BSSs. The variability is from \citealt{Kaluzny_10}:
  SXP means SX Phoenicis, while EW is for contact binary. \#100108 does not have the v sin(i) error because the estimated v sin(i) is a lower limit.}
\end{table*}

\section{Observations}
\label{sec:obs}
This work is based on high resolution stellar spectra acquired with
the multi-object spectrograph FLAMES-GIRAFFE (\citealt{Pasquini_02})
mounted at the Very Large Telescope of European Southern Observatory
(ESO) under program 093.D-0270 (PI: Lovisi). All spectra have been
acquired with the HR2 grating, with a spectral resolution R = 22,700
in the wavelength range $\Delta \lambda = 3854 - 4049 \AA$, in order
to sample the Ca II K line at wavelength $\lambda = 3933.663 \AA$ that is very sensitive to the rotation of stars at high temperature (BSS and HB). For RGBs we used different lines of Fe and Ti. A
total of 16 exposures of 2760 seconds each have been secured. The
observations have been performed in 9 nights between June and July
2014. Six exposures had bad sky conditions (moon distance and airmass)
so we excluded them from the analysis.  The spectra have been acquired
for 115 targets selected along the HB, the RGB, and the BSS sequence
in the UV and optical CMDs presented in \citealt{Lanzoni_07}.  The
standard pre-reduction steps (bias subtraction, flat field correction,
wavelength calibration, and the one-dimensional spectra extraction)
have been performed using the dedicated ESO pipeline \footnote{\url{http://www.eso.org/sci/software/pipelines/}}.  The sky
contribution has been subtracted from the observed spectra by using a
master-sky spectrum obtained as the median of all the sky spectra of
each exposure. Then, the sky-subtracted and heliocentric
velocity-corrected spectra of each target have been combined together. The S/N ratio for the BSS sample ranges between 15 and 30. Instead, the S/N ratio for HBs and RGBs ranges between 25 - 70 and 15 - 70 respectively.

\section{Radial velocities and cluster membership}
\label{sec:rv_membership}
The radial velocities (V$_r$) of the sampled stars have been measured
by using the IRAF task \textit{fxcor} that performs a
cross-correlation between the observed spectrum and a template
spectrum of known V$_r$ (\citealt{TonryDavis_79}). As templates we
adopted synthetic spectra computed with the code SYNTHE
(\citealt{Sbordone_04}; \citealt{Kurucz_05}) for BSSs, HB and RGB stars, to take into account
their different stellar parameters and the different line strengths.
The last version of the Kurucz/Castelli linelist for atomic and
molecular transitions has been adopted. We calculated the model
atmospheres for the synthetic spectra with the ATLAS9 code
(\citealt{Kurucz_93}; \citealt{Sbordone_04}) under the assumptions of
local thermodynamic equilibrium (LTE) and plane-parallel geometry, and
adopting the opacity distribution functions by
\citealt{CastelliKurucz_03} with no inclusion of the approximate
overshooting prescription (\citealt{Castelli_97}).

\begin{figure}
 \centering
 \includegraphics[width=\columnwidth]{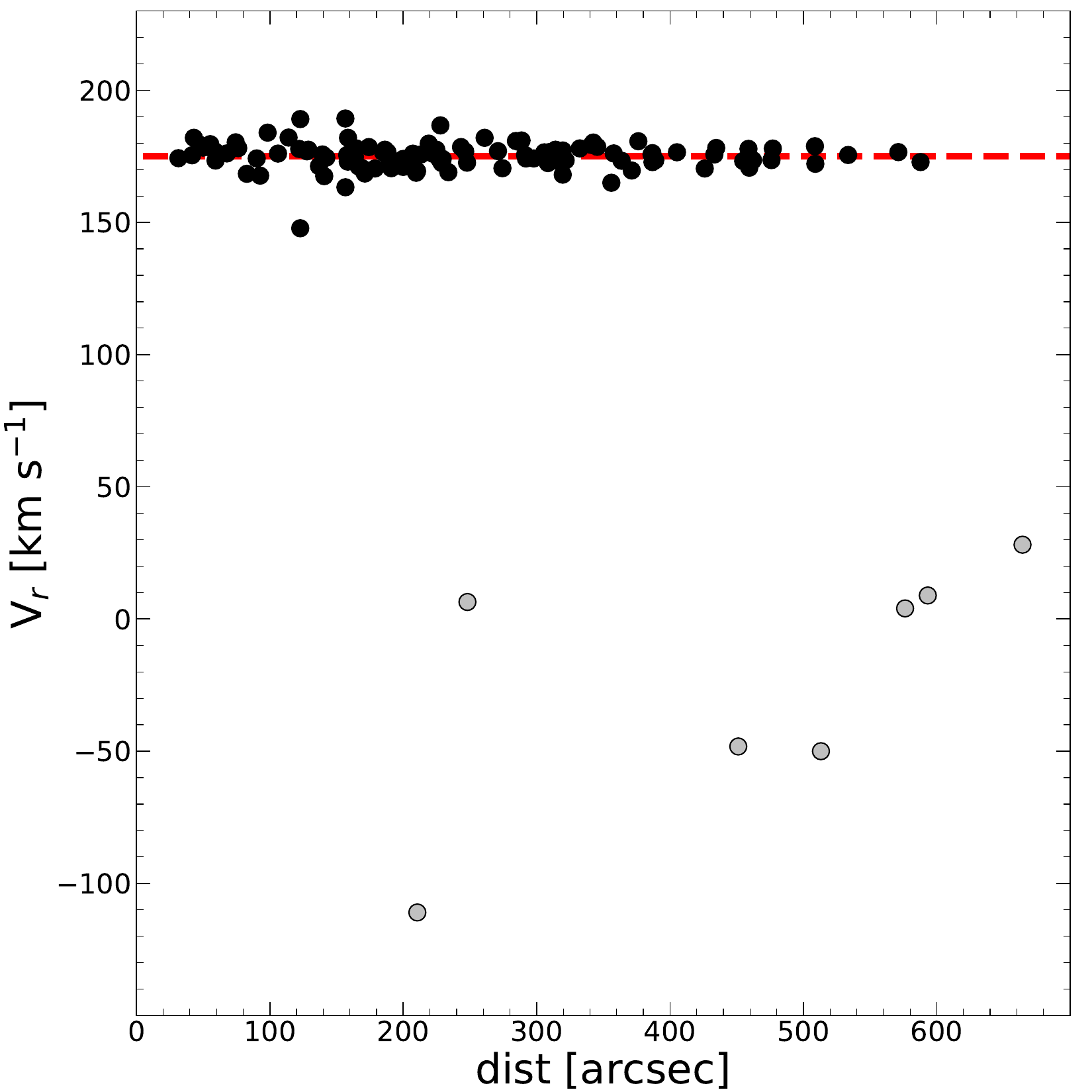}
 \caption{Radial velocity distribution, as a function of the distance
   from the cluster center, obtained from the acquired spectra.  The
   black and gray circles correspond to cluster members and Galactic
   field interlopers, respectively. The red dashed line marks the
   average radial velocity of the member star sample.}
 \label{fig:RV_distribution}
 \end{figure}

Figure \ref{fig:RV_distribution} shows the distribution of the
measured V$_r$ of the 115 targets as a function of the distance from
the cluster center. The vast majority of the targets defines a narrow distribution strongly peaked at the systemic velocity of the cluster. 
The mean value and the standard deviation of the mean obtained from our sample turn out to be 175.1 $\pm$ 0.5 km s$^{-1}$, while the standard deviation of the sample is $\sigma$ = 5.2 km s$^{-1}$, and nicely agrees with the systemic velocity quoted in \citealt{BaumgardtHilker_2018} ($<V_r>$ = 174.8 $\pm$ 0.2 km s$^{-1}$). 
Such a narrow distribution allows a straightforward identification of
the few field stars contaminating the observed sample: indeed only 7
targets (corresponding to $6\%$ of the total sample) display much
smaller radial velocities (well below $<V_r> = 50$ km s$^{-1}$) and can thus
be safely considered as field interlopers (gray circles in
Fig. \ref{fig:RV_distribution}). 
Thus, we considered as member stars those within 3$\sigma$ from the cluster systemic velocity. Only one star (namely \#2700733) is marginally consistent with this assumption within the errors. This is a fast rotator and for this reason the measure of its radial velocity turns out to be more uncertain than the others (see Table 1). However (as shown in Figure 1) the mean radial velocity of field stars in this region of the sky ($<V_r> = -23.1 \pm 18.5$ km s$^{-1}$)  is very different from that of the cluster ($<V_r>=175$ km s$^{-1}$). Hence, the probability that this star is a field interloper is very low. Moreover, its proper motion (from \textit{Gaia} DR3) turns out to be fully consistent with that of the cluster, thus further confirming its membership.
In summary, our sample of likely cluster
members include 32 BSSs, 30 RGB and 46 HB stars. 
The 32 surveyed BSSs are a representative sub-sample of the entire population, which in this cluster counts 55-58 objects. In addition, they have been selected to secure adequate sampling of both the luminosity extension of the BSS sequence, and the BSS radial distribution. Also the HB sample can be considered representative, since it covers the entire distribution in color (temperature) of the HB evolutionary stage. On the other hand, the few observed RGB stars are just meant to provide a reference population of slow rotators.  
The most relevant criterium adopted for the target selection is to avoid contamination of the spectra from scattered light from close neighbors. Hence, all the potential targets having stars of comparable or brighter luminosity within a distance of about 3” (which is more than twice the spectroscopic fiber size) have been excluded from the selection.

For the sake of
illustration, Figure \ref{fig:isocrone} shows their position in the (B,
B$-$I) CMD obtained from ground-based photometry \citep{Stetson_19}.

\section{Atmospheric parameters}
\label{sec:atmospheric_parameters}
\begin{figure}
\centering
\includegraphics[width=\columnwidth]{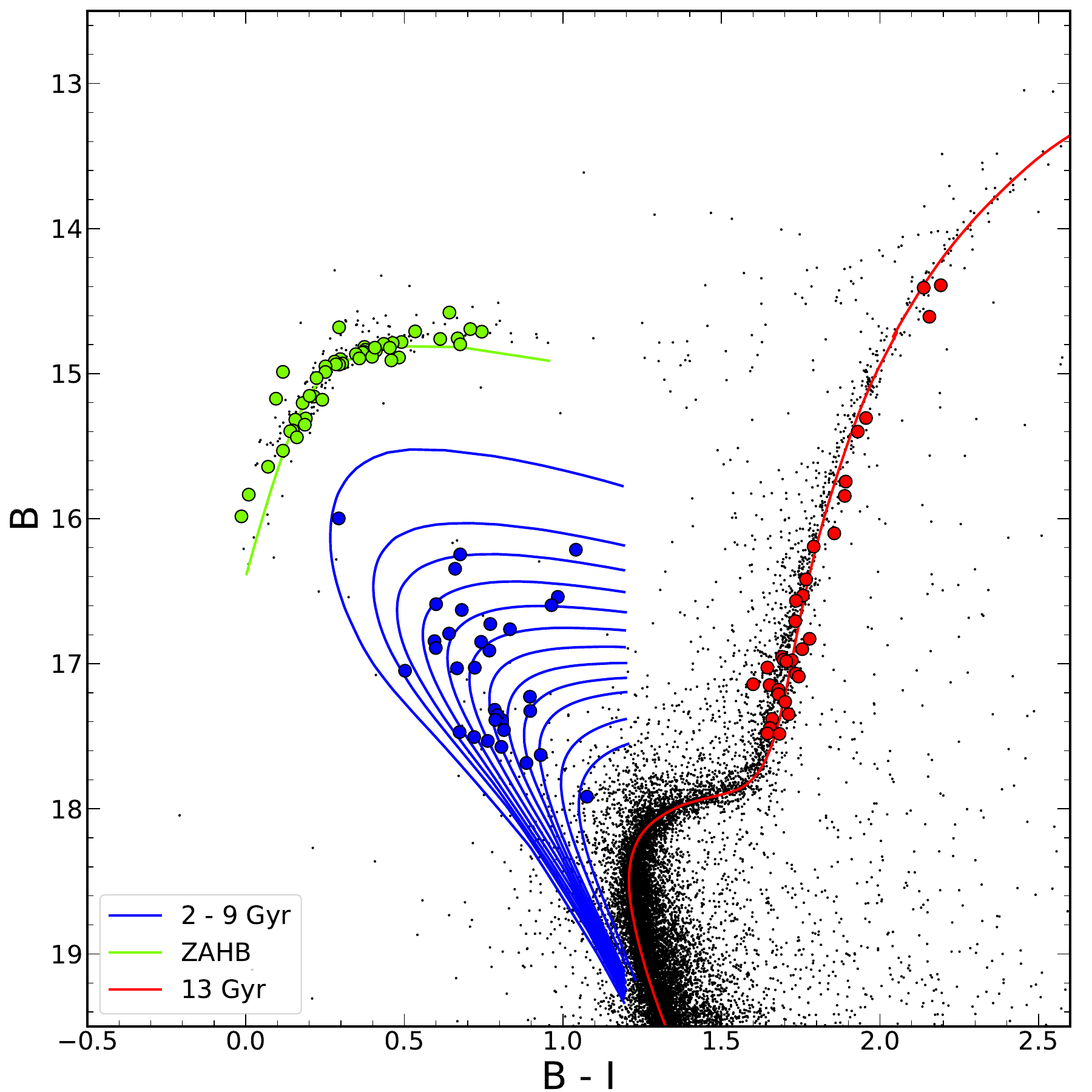}
\caption{CMD of M55 (black dots) with the surveyed BSS, RGB and HB
  stars marked, respectively, as blue, red, and green circles. A set
  of BASTI isochrones (\citealt{Pietrinferni_21}) with ages ranging
  from 2 to 9 Gyr are overplotted as blue lines, a BASTI isochrone of
  13 Gyr is overplotted as red line, and a BASTI HB model is
  overplotted as green line.}
\label{fig:isocrone}
\end{figure}

The values of effective temperature ($T_{\rm eff}$) and surface
gravity (log g) of the spectroscopic targets, have been estimated from
the comparison of their positions in the CMD with reference
theoretical models. We used isochrones and HB models from the
BASTI-IAC database (\citealt{Pietrinferni_21}), adopting an
$\alpha$-enhanced mixture and a metallicity [Fe/H]$ = -1.94$ dex
(\citealt{Harris_96}).  To transform the models into the observational
CMD, we adopted the distance modulus quoted in \citealt{Harris_96} and
then applied a slightly larger value of the reddening, E(B-V) = 0.11
mag instead of E(B-V) = 0.08 (as quoted in \citealt{Harris_96}), to
optimize the match between the models and the data.
We used an HB model with masses ranging from 0.5 M$_{\odot}$ to 0.75
M$_{\odot}$, a 13 Gyr-old isochrone (according to the age estimate
by \citealt{van+13}; 
note that adopting a slightly younger age has a negligible impact on
the temperature and gravity determination) well reproducing the RGB
location, and a set of isochrones with ages ranging from 2 Gyr to 9
Gyr that properly sample the BSS region of the CMD (see solid
lines in Fig. \ref{fig:isocrone}).  Each spectroscopic target has been
orthogonally projected on the closest model, and the corresponding
temperature and gravity have been associated.  The resulting values of
T$_{\rm eff}$ and log g range between 7500-13400 K and 3.0-4.1 dex,
respectively, for the HB stars, while for the RGB targets we find
$\Delta T_{\rm eff} = 4800$-5400 K and $\Delta$ log g = 1.7-3.3 dex,
and for the surveyed BSS sample we obtain $\Delta T_{\rm eff} =
6800$-9700 K and $\Delta$ log g = 3.6-4.3 dex. Finally, a microturbulence of 1 km s$^{-1}$ has been assumed for BSSs
and HB stars, while 1.5 km s$^{-1}$ has been used for the RGB
sample, which are typical values for stars with such temperatures and surface gravities. Note that varying this parameter has a negligible impact on
the derived rotational velocities.

\section{Rotational velocities}
\label{sec:rotational_velocities}

\begin{figure}
\centering
\includegraphics[width=\columnwidth]{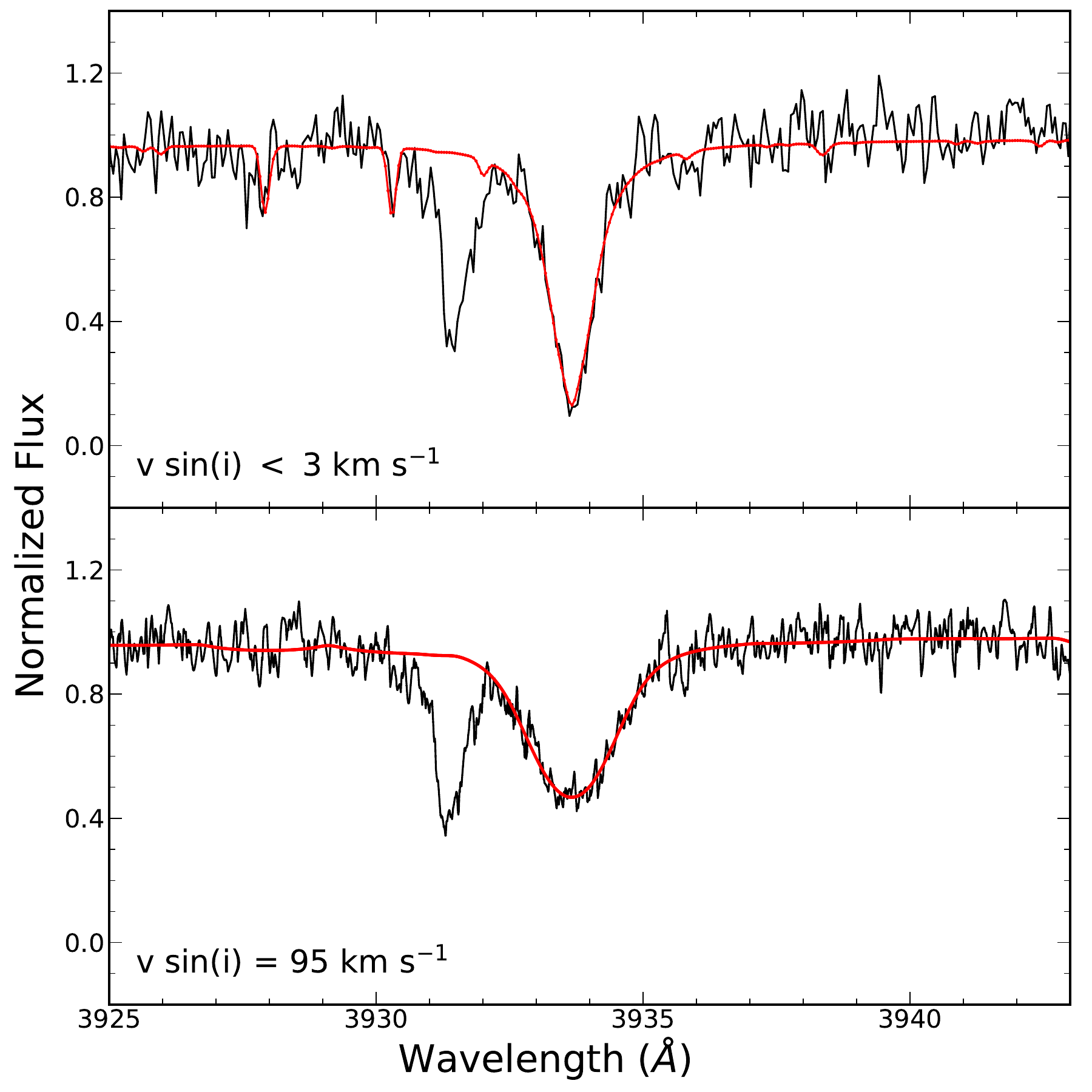}
\caption{Comparison between the observed spectra (black lines) and the
  synthetic spectra (red lines) of two BSSs with different rotational
  velocities (see labels).}
\label{fig:vrot_oss_sint}
\end{figure}

The rotational velocities projected on the line of sight (v sin(i))
have been determined from the comparison between each observed
spectrum and a grid of synthetic spectra computed with the atmospheric
parameters appropriate for the star under consideration, and different
rotational velocities; a $\chi^2$ minimization procedure then finds
the most appropriate value of v sin(i) needed to reproduce the
observations.  This has been done in the spectral region around the Ca
II K absorption line at $\lambda = 3933.663 \AA$.  For the sake of
illustration, Figure \ref{fig:vrot_oss_sint} shows the comparison
between the observed and the synthetic spectra of two BSSs with
different rotational velocities. As can be appreciated, a large value
of v sin(i) causes a large broadening of the line width.\footnote{The
additional absorption line which is visible in the observed spectra at
$\lambda \sim 3931.3 \AA$ is due to the interstellar medium along to
the line of sight. Note that the wavelength of this feature readable
from Figure \ref{fig:vrot_oss_sint} is not the rest frame one, because we
shifted the observed stellar spectrum to the rest frame to make an
appropriate comparison with the synthetic one, but the interstellar
medium has a radial velocity different from that of M55.} For very slowly rotating BSSs we assume an upper limit of 3  km s$^{-1}$  for their v sin(i) value, because this is the typical uncertainty obtained from the Monte Carlo simulations for these objects.
\\
The uncertainties in the rotational velocities obtained with this
procedure have been estimated by using Monte Carlo simulations: a
simulated spectrum with the same signal-to-noise ratio of the observed
one has been obtained by randomly adding poissonian noise to the
best-fitting synthetic template, and its rotational velocity has been
determined with the same procedure applied to the data; after 300
realizations of such realistic simulated spectra, the standard
deviation of the derived v sin(i) distribution has been adopted as
1$\sigma$ uncertainty in the rotational velocity of each star. The
estimated uncertainties range from a few (2-5) km s$^{-1}$ for slowly
rotating stars, up to 10-15 km s$^{-1}$ for the fast-spinning targets.

\begin{figure}
\centering
\includegraphics[width=\columnwidth]{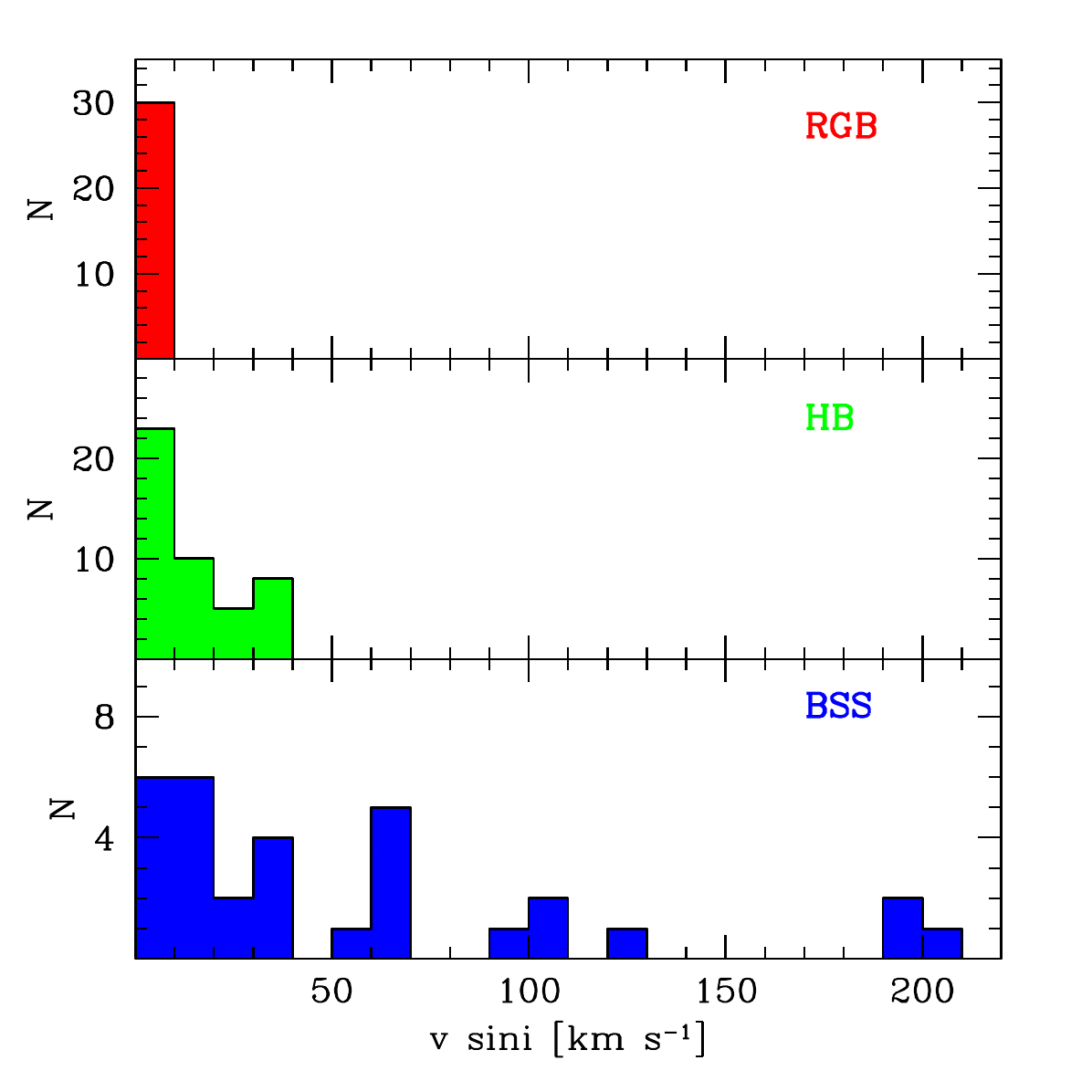}
\caption{Rotational velocity distributions of RGB stars (top panel),
  HB stars (middle panel), and BSSs (bottom panel).}
\label{fig:vsini_M55}
\end{figure}
\begin{figure}
\centering
\includegraphics[width=\columnwidth]{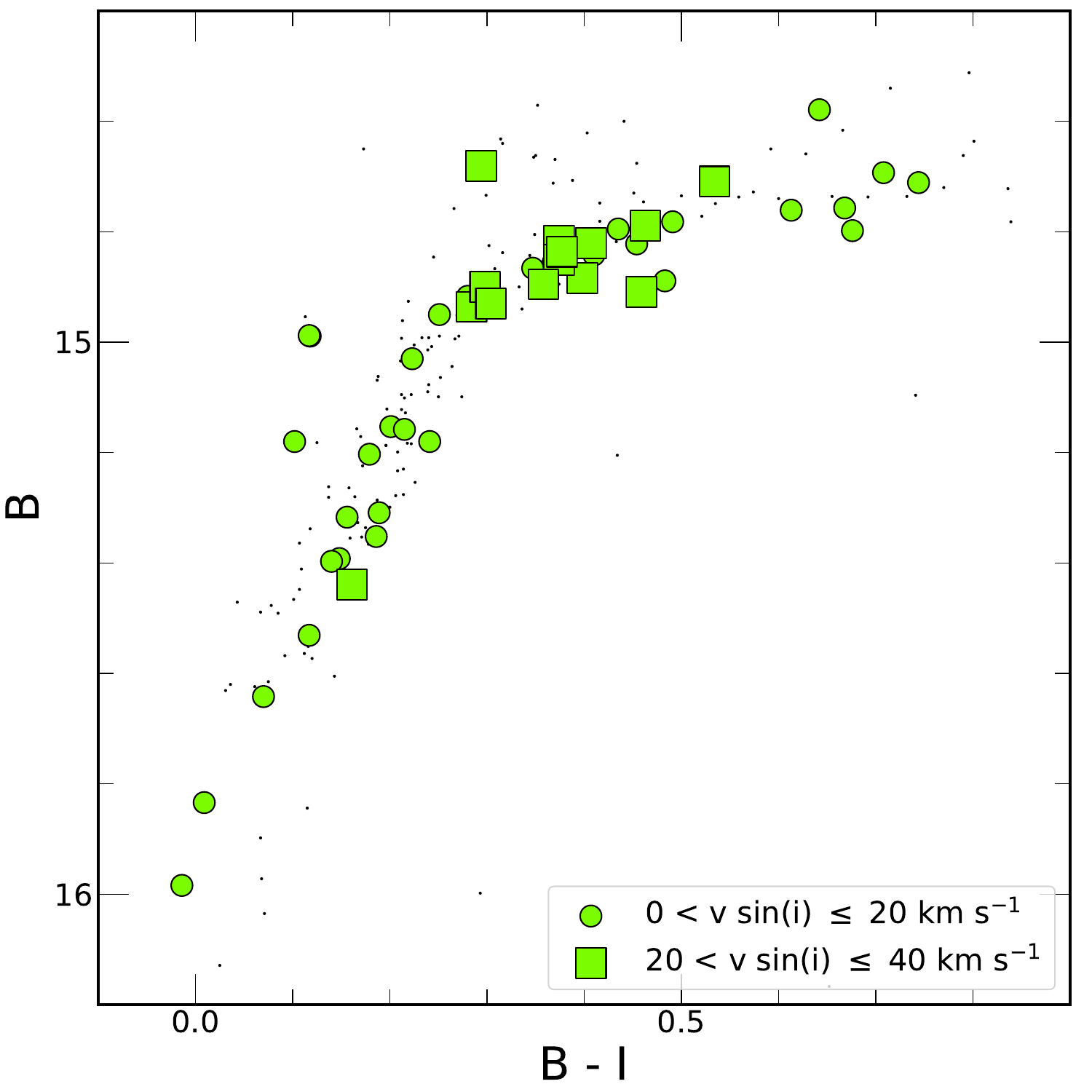}
\caption{HB region of the CMD of M55 with the observed stars
  highlighted with different markers: small green circles for HB stars
  with v sin(i) $\leq 20$ km s$^{-1}$, large green squares for those
  rotating faster (up to 40 km s$^{-1}$).}
\label{fig:vsinihb_cmd}
\end{figure}

Figure \ref{fig:vsini_M55} shows the comparison of the rotational
velocity distributions obtained for the reference populations (RGB and
HB stars) and the BSS sample.
All RGB stars are consistent with null or very slow rotation (v sin(i)
$<$ 10 km s$^{-1}$), while the HB targets show intermediate rotational
velocity values, always smaller than 40 km s$^{-1}$. The distribution
found for BSSs, instead, is much broader, with a tail extending up to
rotational velocities of $\sim 200$ km s$^{-1}$.  Assuming the same
definition of FR stars provided by \citealt{Ferraro_23b} (namely, stars with rotational velocity $\ge 40$ km s$^{-1}$), we conclude that the fraction of FR objects is zero in both the reference (RGB and HB) samples, while it rises to $\sim 47 \pm 14 \%$ (15/32) for the BSS population.  This is the largest fraction of FR BSSs ever detected in a GC. This result will be discussed in Section \ref{sec:discussion}.
\\
Figure \ref{fig:vsinihb_cmd} shows a zoom in the HB region of the CMD,
with different markers corresponding to stars spinning faster and slower than 20 km s$^{-1}$ (see the legend). With the exception of
just one object (out of 13), the HB stars showing large v sin(i) are
located in the redder part of the sequence, at (B$-$I) $>0.2$, while at
bluer colors only slowly rotating HB stars are found.  A decrease of
v sin(i) for increasing effective temperature in HB stars has been
already observed in other GCs, for instance in M13, M15 and NGC 6397
(\citealt{Behr_00a}, \citealt{Behr_00b}, \citealt{Lovisi_12}), but the physical mechanism
that reduces the rotation in hot HB stars is not fully understood
yet. \citealt{SillsPinsonneault_00} suggested that this might be due to
the onset of gravitational settling mechanisms. The strong gradient in
the element abundances can suppress the angular momentum transport and
this prevents the stars to have high v sin(i). \citealt{VinkCassisi_02}
argued that the decrease in rotational velocity for HBs hotter than
11,000 K is due to a loss of angular momentum caused by stellar winds
set up by the radiative levitation.

\begin{figure}
\centering
\includegraphics[width=\columnwidth]{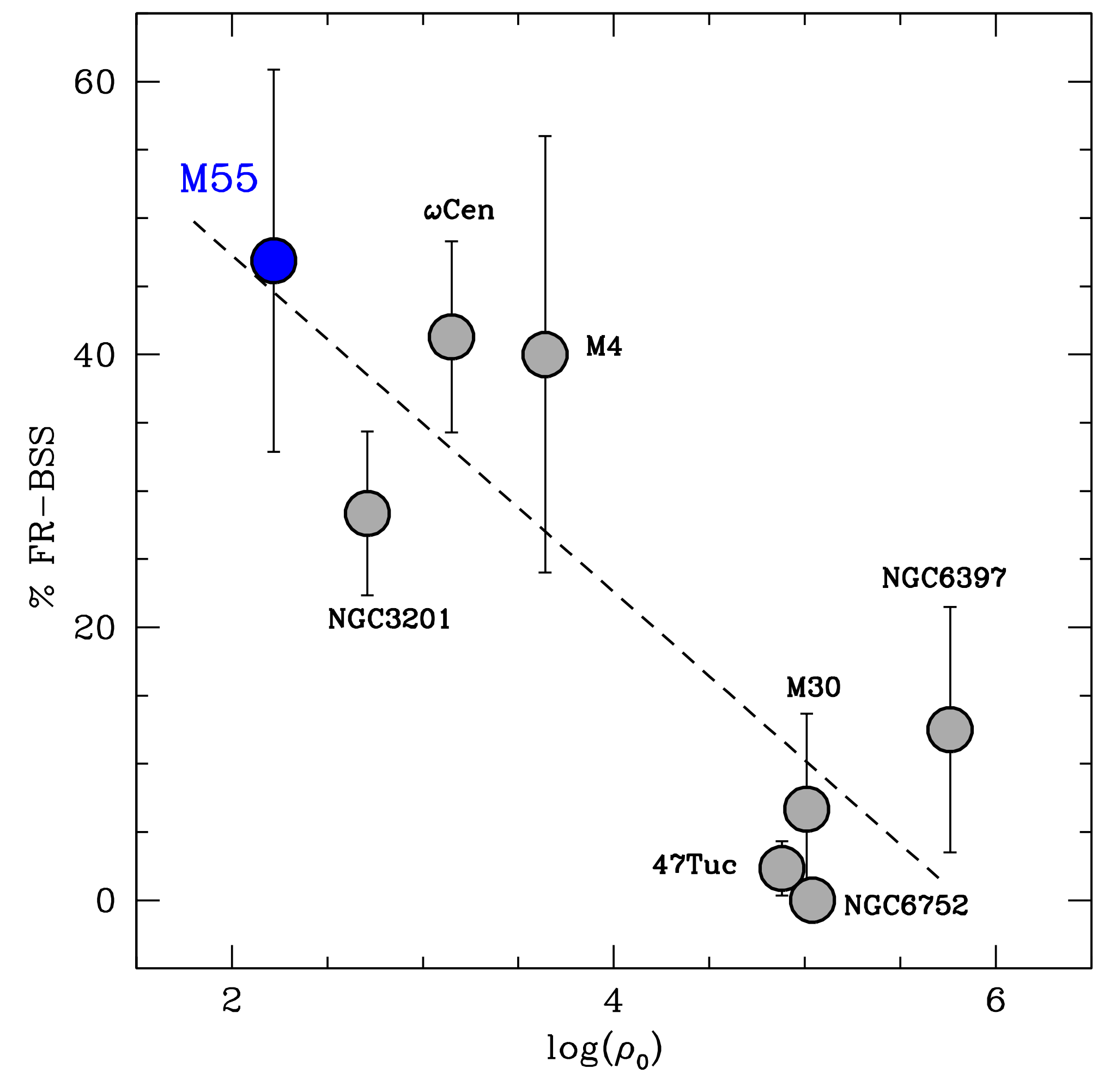}
\caption{Percentage of FR-BSSs as a function of the central density of
  the parent cluster, for the systems studied in
  \citealt{Ferraro_23b}. The position of M55 is highlighted in blue.}
\label{fig:rot}
\end{figure}

\begin{figure}
\centering
\includegraphics[width=\columnwidth]{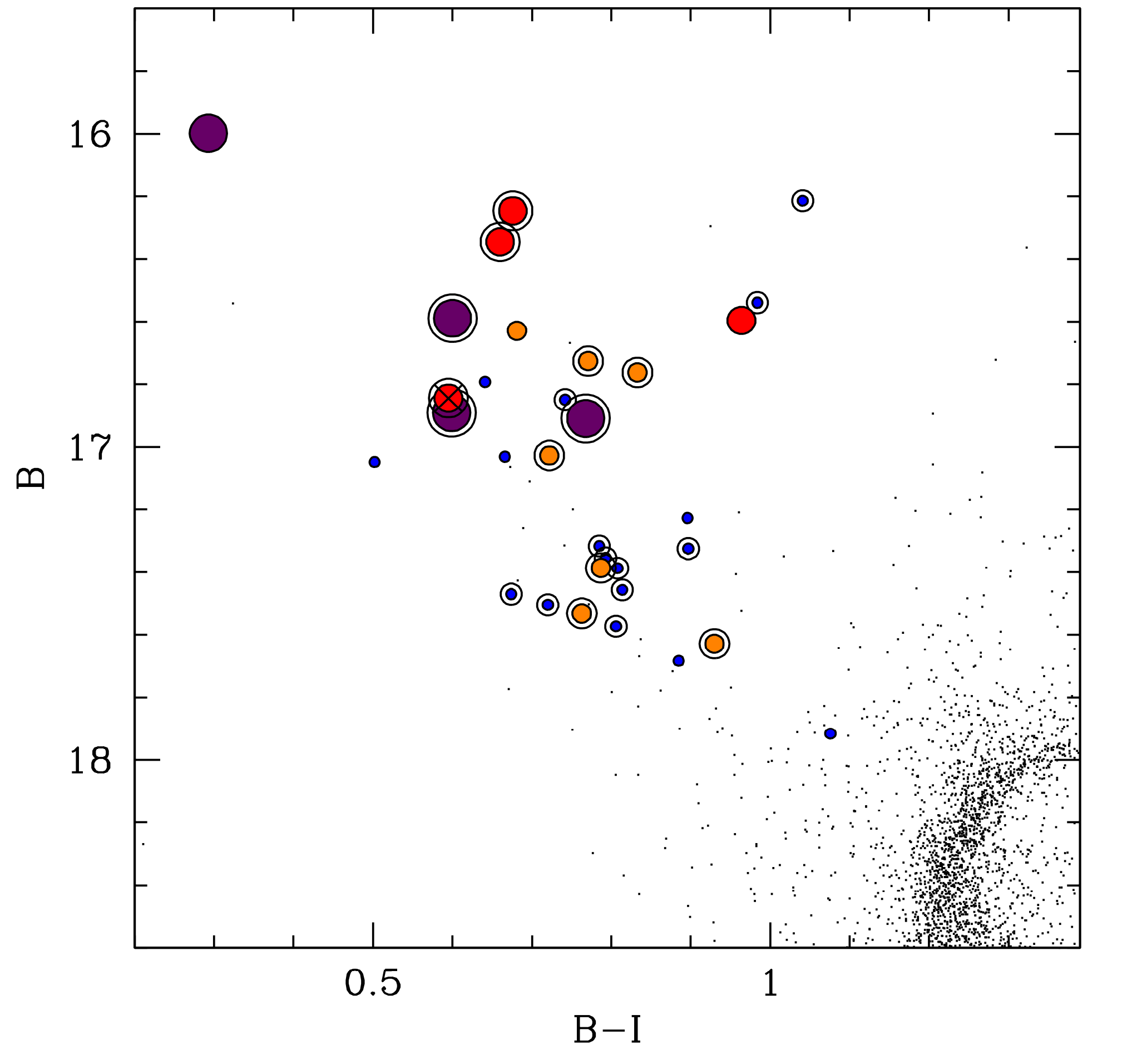}
\caption{CMD of M55 zoomed in the BSS region, with the measured BSSs
  highlighted as large colored circles. The slowly rotating BSSs are
  marked with small blue circles. The FR-BSSs are plotted as
  increasingly larger circles for different colors for increasing
  rotational velocity (in units of km s$^{-1}$): orange color for
  $40\le {\rm v sin(i)} < 80$, red color for $80\le {\rm v sin(i)} <
  180$, and violet for ${\rm v sin(i)} \ge180$ km s$^{-1}$.  All the
  BSSs identified as variable stars in the \citealt{Kaluzny_10}
  catalog are highlighted with large open circles. The FR-BSS
  classified as contact binary is marked with a large black cross.}
\label{fig:var}
\end{figure}

\section{Discussion and conclusions}
\label{sec:discussion}
This work presents the results of a study about the rotational
velocities of BSSs in the Galactic GC M55. As shown in Figure
\ref{fig:vsini_M55}, while the reference (RGB and HB) stars are all
slow rotators, with values of v sin(i) below the threshold of 40
km s$^{-1}$, the BSS rotational velocity distribution shows a more
complex distribution with a long tail extending at high values, up to
$\sim 200$ km s$^{-1}$. This distribution is similar to those found in
M4 (\citealt{Lovisi_10}), $\omega$ Centauri (\citealt{Mucciarelli_14})
and NGC 3201 (\citealt{Billi_23}). All these clusters are part of the
sample studied in \citealt{Ferraro_23b}, where a significant
anti-correlation has been found between the fraction of FR-BSSs and
some structural and dynamical parameters of the parent cluster (such
as the central density and the concentration parameter). In Figure
\ref{fig:rot} we report the behavior of the percentage of FR-BSSs as
a function of the parent cluster central density, with M55 highlighted
in blue. As can be seen, this is the system with the largest fraction
of FR-BSSs and the smallest central density in the sample of GCs
investigated so far.

Figure \ref{fig:var} shows the location of the measured BSSs in the
CMD, plotted as circles of different sizes and colors depending on
their rotational velocity: the slowly rotating BSSs (with v sin(i) $<
40$ km s$^{-1}$) are plotted as small blue circles, while FR-BSSs are
increasingly larger circles with orange color for v sin(i) between 40
and 80 km s$^{-1}$, red color for rotational velocities ranging from
80 to 180 km s$^{-1}$, and violet color for the high-rotation tail,
with v sin(i) $\ge 180$ km s$^{-1}$.  We also checked our BSS sample
for variability, performing a cross-correlation with the catalog of
\citealt{Kaluzny_10} and finding that 23 (out 32) of the sampled BSSs
are indeed variables (see large empty circles in Figure
\ref{fig:var}): 12 of them are FR-BSSs, while the remaining 11 are
slow rotators.  Almost all the detected variables (22 out of 23) are
non binary systems, but pulsating stars classified as SX
Phoenicis. The only exception star \#2700834 (red cross in the
figure), which is cataloged as a contact binary with asymmetric
maxima as consequence of mass transfer.  This is a FR-BSS with v
sin(i) = 104 km s$^{-1}$, thus possibly indicating a system where MT
is still ongoing, hence a proto-BSS in its way of formation.

\begin{figure}
\centering
\includegraphics[width=\columnwidth]{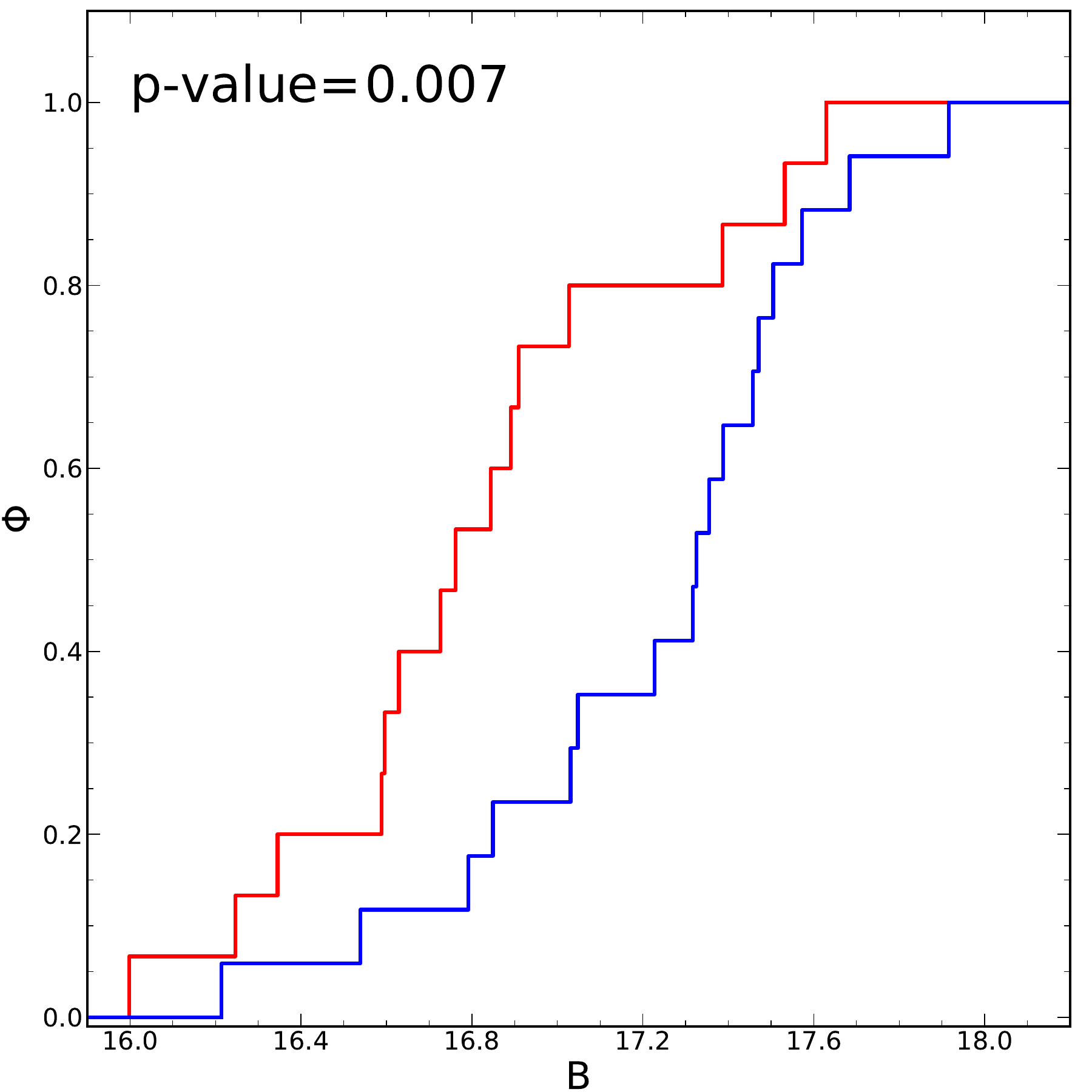}
\caption{Cumulative distribution of FR-BSSs (red line) and slowly
  rotating BSSs (blue line) as a function of the B magnitude.}
\label{fig:cumulativa_M55}
\end{figure}

Indeed, a high rotational velocity is commonly interpreted as a
signature of recent BSS formation activity (see Section
\ref{sec:intro}), but braking mechanisms are expected to then
intervene and slow down the stars (\citealt{LeonardLivio_95};
\citealt{Sills_05}).  Although timescales and efficiencies of these
mechanisms are not fully understood yet, some constraints are starting
to emerge from observational studies.  The analysis of BSS rotation in
open clusters (\citealt{Leiner_18}) suggests a timescale of the order
of 1 Gyr for products of the MT formation scenarios. On other hand, by combining
considerations on the formation epoch of COL-BSSs in the post-core
collapse cluster M30 (\citealt{Ferraro_09,portegies19}) with the BSS
rotation distribution presented in \citealt{Lovisi_13b}, a braking
time-scale of 1-2 Gyr has been inferred also for COL-BSSs
\citep{Ferraro_23b}. Thus, a large percentage of FR-BSSs reasonably
traces a recent activity of BSS formation in the host cluster. The MT scenario should be dominant in low density
environments, where stellar collisions are thought to be negligible.
Hence, the correlation between the percentage of FR-BSSs and the
binary fraction found in \citealt{Ferraro_23b} indicates that an high
percentage of FR-BSSs in low density systems (Fig. \ref{fig:rot})
should be interpreted as the evidence of recent BSS formation activity
due to the MT channel in primordial binaries.

The inspection of the Fig. \ref{fig:var} also suggests that FR-BSSs
tend to populate the brightest portion of the BSS sequence: 12 FR-BSSs
(out of 18 surveyed stars) are found at B $<17.2$, while only 3 FR-BSSs
(out of 14) can be counted in the lower portion of the sequence, thus
indicating that FR-BSSs tend to be more luminous than slowly rotating
BSSs.  This is also supported by the cumulative magnitude distribution
of the samples of slowly and rapidly rotating BSSs shown in Figure
\ref{fig:cumulativa_M55}.  By using a Kolmogorov-Smirnov test, we find
that the probability that the two distributions are extracted from the
same parent population is of the order of 0.7\% (roughly corresponding
to a $2.5\sigma$ significance level).  This is in good agreement with
what observed in NGC 3201 \citep{Billi_23}. Conversely, we found no clear difference between the cumulative color (temperature) distributions of slow and fast rotating BSSs.

\begin{figure}
\centering
\includegraphics[width=\columnwidth]{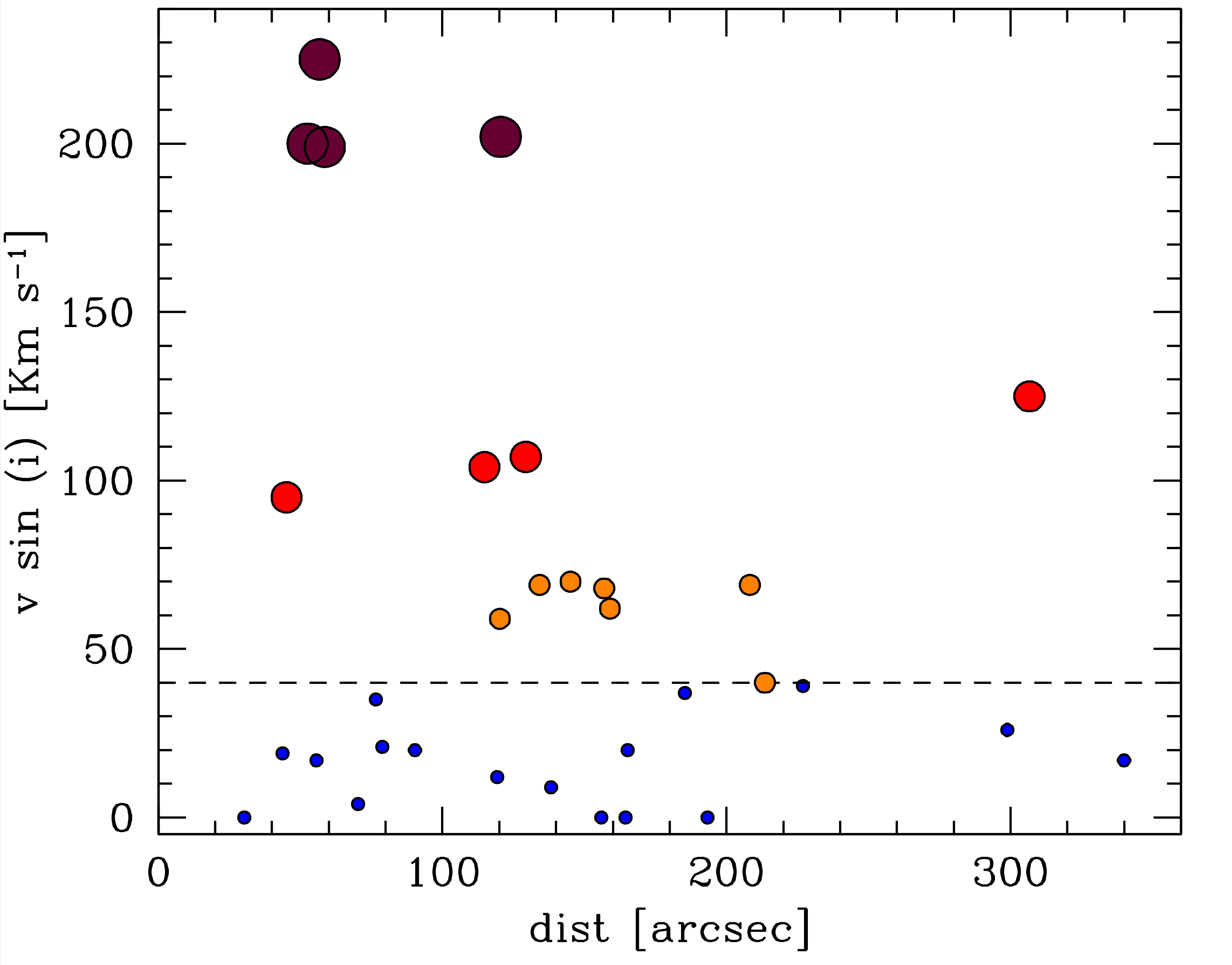}
\caption{BSS rotational velocities as a function of the radial
  distance from the cluster center. The symbols are as in
  Fig. \ref{fig:var}. The dashed line marks the threshold adopted to
  distinguish fast-rotating from slowly rotating stars, namely v
  sin(i) = 40 km s$^{-1}$.}
\label{fig:dist_vsini}
\end{figure}

The dependence between rotational velocity and cluster-centric
distance for the BSSs surveyed in M55 is plotted in Figure
\ref{fig:dist_vsini}.  The trend is less clear here than in NGC 3201
(compare with Fig. 8 in \citealt{Billi_23}), and the
Kolmogorov-Smirnov test detects no significant difference between the
radial distributions of fast and slowly rotating BSSs.  Nevertheless,
we can notice that the BSSs with extremely large rotational velocities
(v sin(i) $>$ 200 km s$^{-1}$), which are also the brightest in the
surveyed sample (see Fig. \ref{fig:var}), are all concentrated in the
inner region of the cluster.  As discussed in \citealt{Billi_23}, this
can be explained by assuming that the fastest (and brightest) BSSs are
also the most massive ones, and they therefore suffered the effect of
mass segregation that made them sink toward the cluster center.  The
difference in the radial distributions observed in M55 and NGC 3201 is
consistent with their different value of the A$_{rh}^+$ parameter,
which quantifies the level of dynamical evolution experienced by the
host cluster \citep{Alessandrini_16}.  According to \citealt{Ferraro_23a} (see also \citealt{Lanzoni_16,Ferraro_18}), M55 has a value of
A$_{rh}^+$ significantly smaller than NGC 3201 (A$_{rh}^+$=0.10,
compared to 0.19). This clearly indicates that M55 is less dynamical
evolved than NGC 3201, consistently with the detected difference in
the trend between v sin(i) and radial distance between the two
clusters.

\begin{figure}
\centering
\includegraphics[width=\columnwidth]{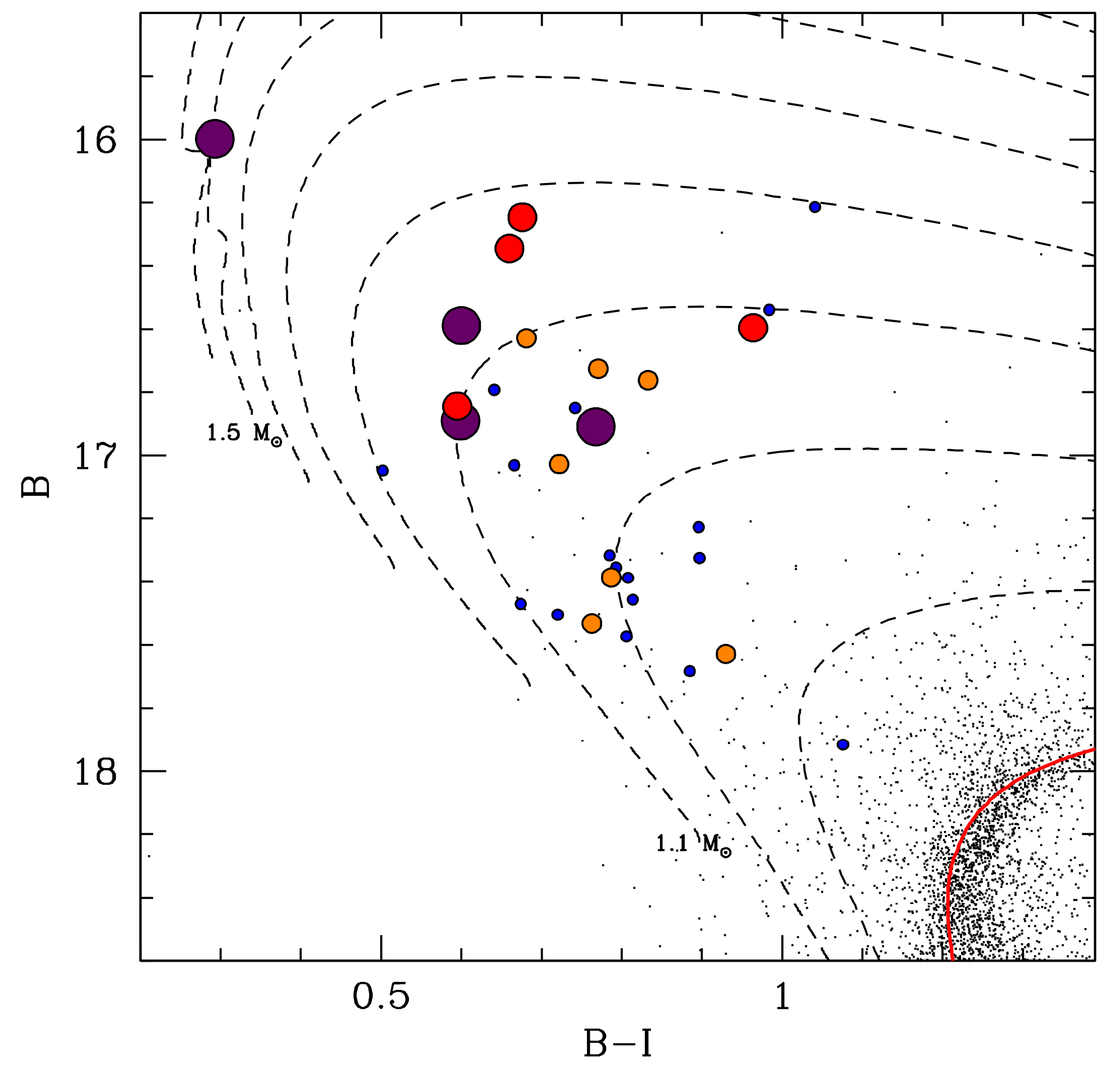}
\caption{CMD of M55 zoomed in the BSS region, with the surveyed BSSs
  highlighted as colored circles (see Figure \ref{fig:var}).  The dashed lines are evolutionary tracks extracted from the BASTI database
  (\citealt{Pietrinferni_21}) for stellar masses ranging from 0.9 to
  $1.6 M_{\odot}$, stepped by $0.1 M_{\odot}$. The red solid line is
  the evolutionary track at $0.8 M_{\odot}$ reproducing the
  cluster MS-TO region.}
\label{fig:bss_ste}
\end{figure}

\begin{figure}
\centering
\includegraphics[width=\columnwidth]{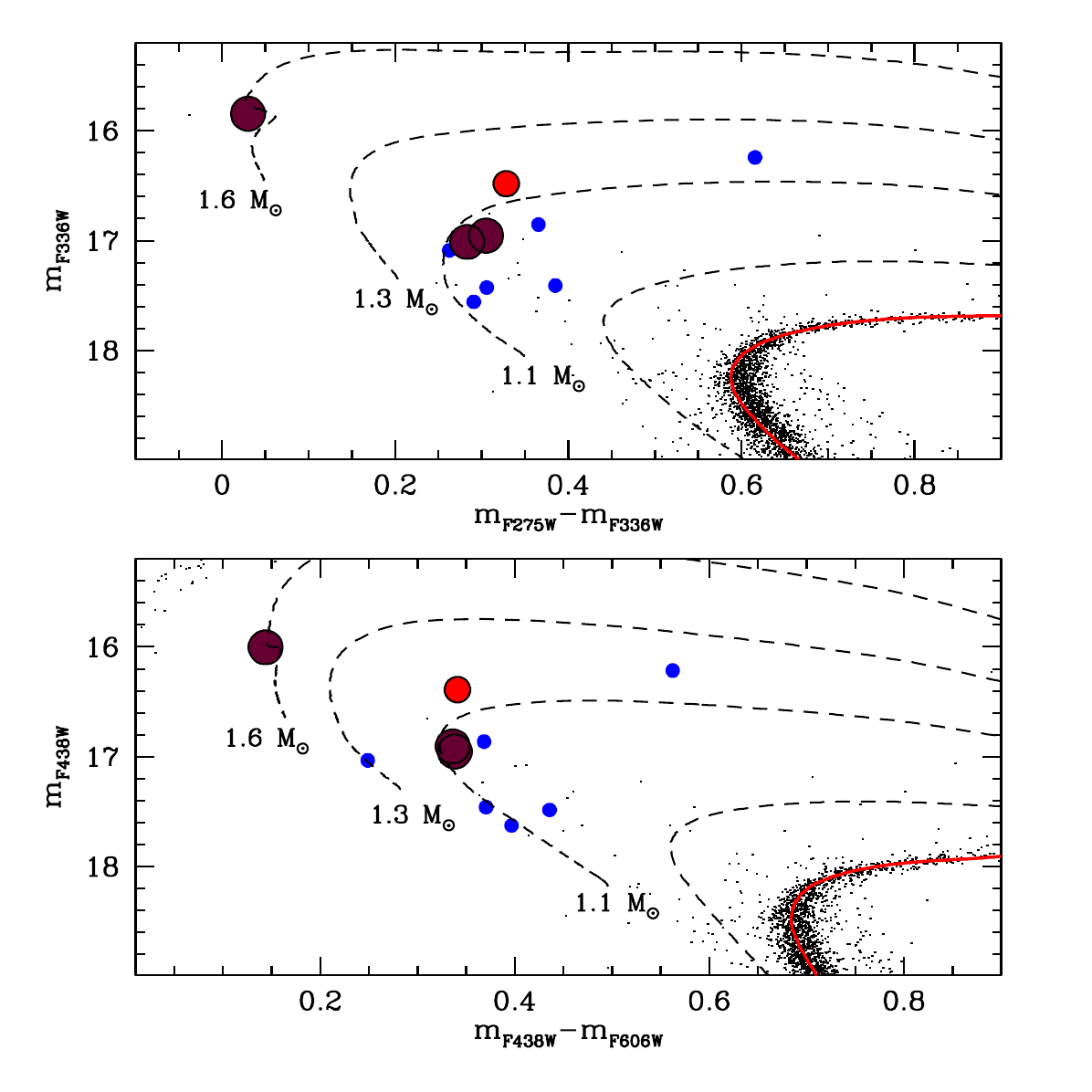}
\caption{As in Fig. \ref{fig:bss_ste}, but for the BSSs sampled in the
  HST Wide Field Camera 3 (WFC3) field of view and for different color-magnitude
  combinations (see x- and y-labels). The four FR-BSSs are \#100108
  (the brightest and bluest one), \#200204 and \#200196 (the other large violet
  circles), and \#200142 (red circle).}
\label{fig:bss_hst}
\end{figure}

The comparison of the CMD position of the measured BSSs with
theoretical evolutionary tracks can provide some additional
insights. We should remind that caution is needed in estimating
the mass of BSSs from the comparison with evolutionary tracks computed
for normal single stars.  However, \citealt{Raso_19} found a reasonable
agreement between the masses derived from single star evolutionary
tracks and those estimated from the fitting of the observed spectral
energy distribution, thus suggesting that the former can be used at
least as first-guess indication.  Figure \ref{fig:bss_ste} shows the
CMD of M55 zoomed in the BSS region, with the dashed lines
corresponding to single-mass theoretical tracks extracted from the
BASTI database \citep{Pietrinferni_21} for masses ranging from 0.9 to
$1.6 M_\odot$, stepped by $0.1 M_\odot$. The position of the fastest
rotators in the sample (i.e., the four BSSs with rotation velocity
larger than 190 km s$^{-1}$) suggests that 3 of them (namely \#200196,
\#200204, \#2700733) have masses of the order of 1.1 $M_\odot$, and
the brightest object (\#100108) is as massive as $\sim 1.5 M_\odot$,
very close to twice the cluster TO-mass.

This latter deserve a few comments. In fact, the Ca II K line is not
visible in the spectrum of this very hot (T$_{\rm eff}>9700$ K) BSS,
because its extremely large rotational velocity has spread out the
line and made it indistinguishable from the continuum. Hence, the
assumed velocity of 200 km s$^{-1}$ is indeed a lower limit.  This BSS
is located in the innermost region of the cluster, at only $52\arcsec$
from the center, where optical, ground-based observations could be
problematic.  To confirm its peculiar position in the CMD, we
therefore used HST data acquired with the WFC3 in the context of the
HST UV Legacy Survey of Galactic GCs (see \citealt{piotto+15,
  Nardiello_18}).  Figure \ref{fig:bss_hst} shows the CMD of M55
obtained in different HST filter combinations, with the BSSs included
in the WFC3 field of view highlighted as large colored circles (same
symbols as in Fig.  \ref{fig:bss_ste}). Clearly, star \#100108 is the
bluest and the brightest BSS, with a position well compatible with a
mass of the order of $1.5-1.6 M_\odot$, in all filter combinations.

The evolutionary time of such a relatively massive star is expected to be quite
fast, thus indicating that the formation of this BSS must have
occurred recently.
Interestingly, the evolutionary time read along the stellar track
indicates that a $1.5 M_{\odot}$ star needs approximately 1 Gyr to
reach the current CMD position of star \#100108, from the instant of
formation.  Thus, this seems to be a BSS that, 1 Gyr after its
formation, is spinning at a velocity larger than 200 km s$^{-1}$.
Following this line of reasoning, the other three very FR-BSSs, which
have masses significantly smaller ($\sim 1.1 M_{\odot}$) than BSS
\#100108, should be still spinning very rapidly after 3 Gyr from their
formation.  The reading of the evolutionary time from single star
theoretical tracks therefore suggests a not particularly efficient
braking mechanism, with newly-formed BSSs being able to maintain a
strong rotation for a few Gyrs.  However, we notice that also the
FR-BSS \#2700834 (which is classified as a contact binary with ongoing
MT activity; see \citealt{Kaluzny_10}) is located at the same position
in the CMD (see Fig. \ref{fig:var}). This suggests that 1-3 Gyr old
BSSs, and BSSs on the way of their formation can occupy the same
region of the diagram during their evolution, as it is indeed
predicted by theoretical MT models specifically designed to describe
the BSS population of M30 (see \citealt{xin+15,jiang+17}).  Thus,
while a first-guess estimate of BSS masses can be reasonably derived
from the comparison with single star evolutionary tracks, extreme
caution must be taken in reading the evolutionary time-scale from the
same models. This emphasizes the urgency of complete grids of MT-BSS
models for an appropriate reading of the BSS evolutionary path.
  
\begin{acknowledgements}
This work is part of the project {\it Cosmic-Lab} (Globular Clusters
as Cosmic Laboratories) at the Physics and Astronomy Department
``A. Righi'' of the Bologna University
(http://www.cosmic-lab.eu/Cosmic-Lab/Home.html). A.B. acknowledges
funding from the European Union NextGenerationEU.
 
\end{acknowledgements}

%
\bibliographystyle{aa} 
\bibliography{biblio} 
%


\end{document}